\documentclass{aa}
\usepackage{graphicx}
\usepackage{txfonts}
\usepackage[authoryear]{natbib}
\bibpunct{(}{)}{;}{a}{}{,}

\newcommand{\xte}{{\it RXTE}}

\def\simless{\mathbin{\lower 3pt\hbox
     {$\rlap{\raise 5pt\hbox{$\char'074$}}\mathchar"7218$}}}   
\def\simmore{\mathbin{\lower 3pt\hbox
     {$\rlap{\raise 5pt\hbox{$\char'076$}}\mathchar"7218$}}}   

\def\la{~\raise.5ex\hbox{$<$}\kern-.8em\lower 1mm\hbox{$\sim$}~}
\def\ma{~\raise.5ex\hbox{$>$}\kern-.8em\lower 1mm\hbox{$\sim$}~}

\begin{document}
\title{A jet model for Galactic black-hole X-ray sources: 
Some constraining correlations}

\author{
N. D. Kylafis\inst{1,2} \and I. E. Papadakis\inst{1,2} \and P. Reig\inst{2,1}
\and D. Giannios\inst{3} \and G. G. Pooley\inst{4} 
}
\offprints{N. D. Kylafis;  e-mail: kylafis@physics.uoc.gr}
\institute{
Physics Department, University of Crete, P.O. Box 2208,
   710 03 Heraklion, Crete, Greece 
\and 
Foundation for Research and Technology-Hellas, P.O. Box 1527, 
   711 10 Heraklion, Crete, Greece
\and
Max Planck Institute for Astrophysics, Box 1317, 85741 Garching, Germany
\and
Astrophysics Group, Cavendish Laboratory, Madingley Road, Cambridge CB3
OHE, UK
}
 
\date{Received ?/ Accepted ?} 
\abstract
{Some recent observational results impose significant constraints on 
all the models that have been proposed to explain the Galactic black-hole X-ray 
sources in the hard state.  In particular, it has been found that during 
the hard state of Cyg X-1 the power-law photon number 
spectral index, $\Gamma$, is 
correlated with the average time lag, $<t_{\rm lag}>$, between hard and 
soft X-rays. Furthermore, the peak frequencies of the four Lorentzians that
fit the observed power spectra are correlated with both $\Gamma$ 
and $<t_{\rm lag}>$. }
{We have investigated whether our jet model can reproduce
these correlations.}
{We performed Monte Carlo simulations of Compton upscattering of
soft, accretion-disk photons in the jet
and computed the time lag between hard and soft photons and the power-law
index $\Gamma$ of the resulting photon number spectra.}
{We demonstrate that our jet model 
naturally explains the above correlations, with no additional
requirements and no additional parameters.
}
{}

\keywords{accretion, accretion disks -- black hole physics -- radiation
mechanisms: non-thermal -- methods: statistical -- X-rays: stars
 }
\titlerunning{Jet model for black-hole sources} 
\authorrunning{Kylafis et al.}
\maketitle
   
\section{Introduction}

The states of Galactic black-hole X-ray sources (GBHs) are determined from the
following quantities:  i) their disk fraction, which is the ratio of the disk
flux to the total flux, both unabsorbed, at $2-20$ keV; ii) the photon number
spectral index, $\Gamma$, of the  hard band,  power-law component at energies
below any break or cutoff;  iii) the root-mean-square (rms) power in the power
spectral density (PSD) integrated from 0.1--10 Hz;  iv) the integrated rms
amplitude of any quasi-periodic oscillation (QPO) detected in the range 0.1--10
Hz \citep{remi06}.  In particular, the so-called hard state is characterized by
a power-law X-ray spectrum with photon  number spectral index $1.5 \simless
\Gamma \simless 2.1$, which suffers an exponential cutoff at $\sim$ 100  or few
100 keV, and contributes more than 80\% of the 2--20 keV flux.  

Several models have succeeded in reproducing the observed X-ray spectrum in this
state \citep[see e.g.][]{tita94,esin97,pout99,reig03,gian04}. When a GBH is in
the hard state, radio emission  is also detected and a jet is either seen or
inferred \citep{fend01}. Again, several models are successful in reproducing the
energy spectrum from the radio domain to the hard X-rays \citep[see
e.g.][]{mark01,vada01,corb02,mark03,gian05}.  The multiplicity of models that
can fit the time average spectrum of GBHs well indicates that this alone is  not
enough to distinguish the most realistic among them.

Timing data of the observed intensity in various energy bands provide a totally
different ``dimension" than the spectral one, and can in principle provide
information on the location, size, physical conditions and kinematics of the
plasma responsible for the X-ray emission in these objects. For example, the
X-ray light curves of GBHs  in different energy bands show similar variations,
but those in the higher energy band generally lag the variations detected in the
lower energy band  \citep{nowa99,ford99}. The time delay, $t_{\rm lag}$, usually
depends  on frequency, $\nu$.  For Cyg X-1  the time delays decrease with
increasing frequency roughly as  $t_{\rm lag} \propto \nu^{-\beta}$, where
$\beta \approx 0.7$. As constraining as it is, this relation has also been
explained in more than one way  \citep[see e.g.][]{pout99,reig03,kord04}.  

Even more constraining to models is the fact that the high-frequency power
spectrum flattens as the photon energy increases \citep{nowa99}. Equivalent
to this is the fact that the width of the  temporal autocorrelation
function of the light curves of Cyg X-1  decreases with increasing photon
energy \citep{macc00}. These observational facts are contrary to what one
expects from simple Comptonization models, yet possible  explanations for
them  have also been offered \citep{koto01,gian04}. 

Significant advances in our understanding of how the X-ray mechanism in
GBHs (and compact accreting objects in general) works  can be achieved when
we combine variability and spectral information. This has been possible in
the last few years, due mainly to \xte.  For example, in the case of Cyg
X-1, \citet{pott03}  used 130 \xte\ observations between 1998 and 2001 to
study  the long-term evolution of the source's power and energy spectrum in
detail. They used 512 s long light curves to estimate the  power spectral
density (PSD), which they fitted with a model that consisted of the sum of
multiple Lorentzian profiles. They generally achieved good fits using four
broad Lorentzian components with typical peak frequencies of $\nu_1\sim
0.2$ Hz, $\nu_2\sim 2$ Hz, $\nu_3\sim 6$ Hz, and $\nu_4\sim 40$ Hz, which
vary in unison with time: their ratios remain constant despite 
each frequency varies by up to a factor of $\sim 5$. Up to now, no
physical explanation of these Lorentzians has been offered. As for the
energy spectrum, they find that a simple model consisting of a power-law
component of index $\Gamma$, a multi-temperature disk black body and a
reflection from neutral material component could fit the energy
spectrum of the source below 20 keV. 

They present convincing evidence that the timing and spectral properties of the
source are closely linked. In fact, a number of the correlations they present
impose stringent constraints on all the  models proposed so far.
These correlations are:

a) The observed photon number spectral index, $\Gamma$, correlates with the
observed average time lag, $<t_{\rm lag}>$, between 2--4 keV and 8--13 keV,
averaged over the 3.2--10  Hz frequency band, in the sense that  $<t_{\rm lag}>$
increases as the spectrum steepens. Therefore, a model that can successfully
explain the spectral slope $\Gamma$ should also predict the right amount of time
delay between these specific energy bands and vice versa: a model that can fit
well  the time lag vs. Fourier frequency relation, should also predict the
appropriate value of the index $\Gamma$  for a given time delay.

b) The observed photon number spectral index, $\Gamma$, also  correlates with
the Lorentzian peak frequencies: the frequencies increase as the spectrum
steepens \citep[see also][]{shap06,shap07}.

In summary, the \citet{pott03} results show that a specific value of  $\Gamma$,
say $\Gamma = 2$, which is relatively easy to obtain with a simple
Comptonization model, must also be associated with {\it specific} values of the
Lorentzian peak frequencies and a {\it specific} value of  $<t_{\rm lag}>$!
These are challenging relations that  any model must be able to explain. In
this work we investigate whether the jet model  that we have proposed can meet
these constraints under reasonable assumptions. 

We think that the jets in GBHs  play a central role in all the observed
phenomena,  not only in the radio emission. We have shown that Compton
upscattering in the jet  of soft photons from the accretion disk can explain: 1)
the X-ray spectrum and the time-lag dependence on Fourier frequency 
\citep[][hereafter Paper I]{reig03}; 2) the narrowing of the autocorrelation
function and the increase of the  rms amplitude of variability with increasing
photon energy  \citep[][hereafter Paper II]{gian04}; 3) the energy spectrum from
radio to hard X-rays of XTE J1118+480, the only source for which we have
simultaneous observations for all energy bands \citep[][hereafter Paper
III]{gian05}.

In this paper we show that the jet model proposed and developed in the
above work can explain the $\Gamma$ -- $<t_{\rm lag}>$ relation of
\citet{pott03}, with no additional parameters or requirements,  if we let
just two model parameters, namely the radius at the base of the jet and the
optical depth along its axis, vary over a reasonable range of values. At
the same time, the model can also explain the "$\Gamma$ -- Lorentzian peak
frequencies $\nu_i$"  relation  qualitatively if we identify $1/\nu_i$, 
$i=1, 2, 3, 4$, with quantities proportional to the characteristic
timescale $M_{\rm out}/{\dot M}_{\rm out}$,  where ${\dot M}_{\rm out}$ is
the rate of mass outflow in the jet and $M_{\rm out}$  the available
mass to power the jet. Finally, the model parameter variations that are
needed to explain the  \citet{pott03} relations also predict that, on
long time scales,  the radio variations should also correlate with
$\Gamma$,  which, as we show, is indeed the case in Cyg X-1. 

In Sect. 2 we describe briefly our model, in Sect. 3 we present and discuss 
our results, and we close with our conclusions in Sect. 4.

\section{The model}

The jet model used here is the same as the one described in Papers I, II
and III. A summary is given below so that the reader becomes familiar with
the  parameters involved. First, we discuss the asumptions made
in the model; then we describe how our Monte Carlo code works and finally
we explain the meaning of the various parameters involved.

\subsection{Description of the jet}

We assume that the jet is accelerated close to its launching area and maintains
a constant (mildly  relativistic) bulk speed, $v_{\parallel} $, thereafter. 
The electron-density profile in the jet is taken to have the form

$$
n_{\rm e}(z) = n_0 (z_0/z),
\eqno(1)
$$ 
where $z$ is the distance from the black hole perpendicular to the  accretion
disk. The parameters $n_0$ and $z_0$ are the electron density and the height of
the base of the jet. The Thomson optical depth along the axis of
the jet is

$$
{\tau}_{\parallel}=n_0{\sigma}_T z_0 \ln (H/z_0),
\eqno(2)
$$
where $H$ is the extent of the jet along the $z$ axis.

Let $\pi R^2(z)$ be the cross-sectional area of the jet as a function of height
$z$. Then from the continuity equation  ${\dot M}_{\rm out}=\pi
R^2(z)m_pn_e(z)v_{\parallel}$,  we obtain the radial extent $R$ of the jet as a
function of height $z$

$$
R(z)=R_0(z/z_0)^{1/2}.
\eqno(3)    
$$
Here $R_0=({\dot M}_{\rm out}/\pi m_pn_0v_{\parallel})^{1/2}$  is the radius of
the jet at its base, ${\dot M}_{\rm out}$  is the mass-outflow rate, $m_p$ is
the proton mass, and $v_{\parallel}$ is the component of the velocity of the
electrons  parallel to the $z$ axis.

For computational  simplicity, we assume that a uniform magnetic field exists
along the  axis of the jet (taken to be the $z$ axis). The $z$-dependence of
the magnetic field is dictated by magnetic flux conservation along the jet axis
to be

$$ B(z)=B_0(z_0/z), \eqno(4) $$ where $B_0$ is the magnetic field strength
at the base of the jet. Also for computational simplicity we take the
electrons' velocity as having two components one parallel
($v_{\parallel}$)  and one perpendicular  ($v_{\perp}$) to the magnetic
field.  The magnitude and direction of $v_{\parallel}$ are fixed; so is the
magnitude of $v_{\perp}$.  However, the direction of $v_{\perp}$ is
continuously changing, being  always tangent to a circle in the
perpendicular plane with respect to the  magnetic field.  

It has been shown (Paper III) that a power-law distribution of electron 
velocities in the rest frame of the flow gives nearly  identical results.  This
is because the dominant contribution to the scatterings comes from the electrons
that have the  lowest velocity (or the  lowest Lorentz $\gamma$ factor), due to
the steep power law of electron  $\gamma$'s required to explain  the overall
spectrum.

\subsection{The Monte Carlo simulations}

For our Monte Carlo simulations we used the same code as in Paper I.  The code
has been checked with an independent code that was used in Papers II and III. 
The procedure we follow is described briefly below.

Photons from the inner part of the accretion disk, in the form of  a blackbody
distribution of characteristic temperature $T_{\rm bb}$, are injected at the
base of the jet with an upward isotropic distribution. Each photon is given a
weight equal to unity (equivalently, beam of flux unity)
when it leaves the accretion  disk, and its time of
flight is set equal to zero.  The optical depth $\tau (\hat n) $ along the
photon's direction $\hat n$ is computed and a  fraction $e^{- \tau (\hat n)}$
of the photon's weight escapes and is recorded.  The rest of the weight of the
photon gets scattered in the jet. If the effective optical depth in the jet is
significant (i.e.,  $\simmore 1$), then a progressively smaller and smaller
weight of the photon experiences more and more scatterings.  When the remaining
weight in a photon becomes less than a small number (typically $10^{-8}$), we
start  with a new photon.  

The time of flight of a random walking photon (or more accurately of its
remaining weight) gets updated at every scattering by adding the last distance
traveled divided by the speed of light.  For the escaping weight along a travel
direction, we add an extra time of flight {\it outside} the jet in order to bring
in step all the photons (or fractions of them) that escape in a given direction
from different points on the  ``surface'' of the jet.  The  longer a
fractional photon stays in the jet, the more energy it gains on average from the
circular motion (i.e., $\vec{v_{\perp}}$) of the electrons.  Such Comptonization
can occur everywhere in the jet.  Yet, a photon that random walks high up in the
jet has a longer time of flight than a photon that random walks near the bottom
of the jet. The optical depth to  electron scattering, the energy shift, and the
new direction of the  photons after scattering are computed using the
corresponding  relativistic expressions.

If the defining  parameters of a photon (position,  direction, energy,
weight, and time of flight) at each stage of its flight are computed, then
we can determine the spectrum of the radiation  emerging from the 
scattering medium and the time of flight of each escaping fractional
photon.  The time of flight of all escaping fractional photons is recorded 
in 8192
time bins of  1/128 s each. In this way, we can compute the number of
photons that are emitted by the jet in each time bin and for any energy band.
In other words, we can create 64 s long light curves for any energy band,
which correspond to the resulting emission of the jet in response to an
instantaneous burst of soft photons that we assume enter the jet
simultaneously.  

Having created light curves in various energy bands, we can now compute
delays between any pair of bands. In our case, we consider the light curves
of the energy bands 2--4 and 8--14 keV (identical to the energy bands of
the light curves that Pottschmidt et al. 2003 used).   Following
\citet{vaug97}, we compute the phase lag and, through it, the time lag
between the two energy bands as a function of Fourier frequency (same as in
Paper I).   Then we compute the average time lag, $<t_{\rm lag}>$,  in the 
Fourier frequency range 3.2--10 Hz (like Pottschmidt et al. 2003).

\subsection{Parameter values}

Our jet model has a number of free parameters, which are the same as in Paper
I: the temperature $T_{bb}$ of the soft photon input,  the extent $H$ of the
jet, the radius $R_0$ and the height $z_0$ of the base of the jet,  the Thomson
optical depth ${\tau}_{\parallel}$ along the axis of  the jet, the parallel
component of the velocity  $v_{\parallel}$ of the electrons,  and the
perpendicular velocity component $v_{\perp}$.

We note here that, due to the mildly relativistic flow in the jet, the
emerging X-ray spectra from the jet depend on the angle between  the line
of sight and the magnetic field direction.    In order to have good
statistics in our Monte Carlo results, we combined all the escaping photons
with directional cosine with respect to the axis of  the jet in the range
$0.2< \cos \theta < 0.6$.  Strictly speaking, this is inconsistent with our
assumption of a uniform  magnetic field.  If the magnetic field and the
bulk flow were uniform,  then only one value  of $\cos \theta$ would be
appropriate for each source.   As discussed in Paper II, a range of
directional cosines is absolutely  necessary for reproducing the
observed phenomena.  Therefore, a  non-uniform (probably a spiraling)
magnetic field permeates the jet  and this, together with the bulk flow,
result in the above range of  directions.  A range of directions is thus
important, but the actual values  of this range are not.

The values of the model parameters are chosen so that all the results of
Papers I, II, and III (apart from the radio emission) are reproduced, and at
the same time the hard band power-law component has a slope of $\Gamma=2.1$
while the average time lag, in the range 3.2--10 Hz, between the 2--4 keV
and 8--14 keV bands is 5 ms (as observed by Pottschmidt et al. 2003). We
call these values the {\it reference values} and they are: $kT_{\rm bb}
= 0.2$ keV, $H=10^5 r_g$,  $R_0=50 r_g$, $z_0=5r_g$, $\tau_{\parallel}=3$,
$v_{\parallel}=0.8 c$,  and $v_{\perp}=0.4 c$,  where $r_g = GM/c^2 = 1.5
\times 10^6$ cm stands for the gravitational radius of a 10 solar-mass
black hole.    

\begin{figure}
\resizebox{\hsize}{!}{\includegraphics{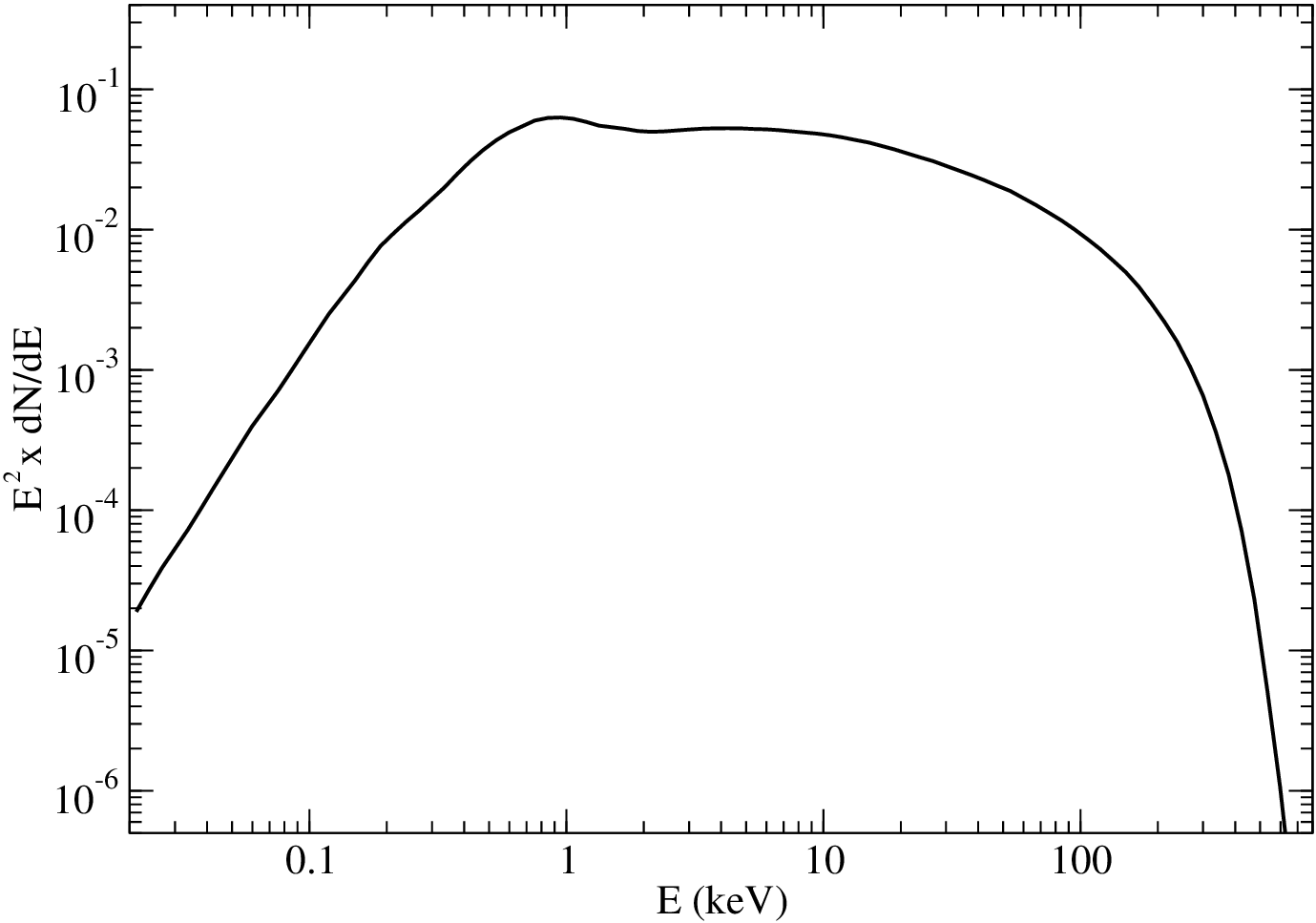}}
\caption[]{Emergent photon-number spectrum multiplied by $E^2$
from our jet model for the reference values of the
parameters given in Sect. 3. The soft X-ray excess below $\sim$ 2 keV is 
essentially the unscattered soft-photon input. When fitted with a power law 
with an exponential cutoff (see Eq. 5), 
we find $\Gamma=2.1$ and $E_{\rm cut}=70$ keV.}
\label{refspec}
\end{figure}

In Fig.~\ref{refspec} we show the emerging X-ray spectrum for the reference
values of the parameters. The spectrum above $\sim 2$ keV can be well-fitted
with a power law with an exponential cutoff

$$
dN/dE\propto E^{-\Gamma}\rm e^{-E/E_{\rm cut}},
\eqno(5)
$$

\noindent where $\Gamma =2.1$ and $E_{\rm cut}=70$ keV. One can also see
the soft X-ray spectrum (below $\sim 2$ keV), which is essentially the
unscattered part of the soft-photon input.

\section{Results and discussion}

\subsection{The spectral index -- time lag relation} 

Our first aim is to investigate whether we can reproduce the $\Gamma$ - 
$<t_{\rm lag}>$ correlation of \citet[][ see their Fig. 7c]{pott03} by
allowing only a few model parameter values to change by a
``reasonable amount" around their reference values. 

It was shown in Paper II that keeping the rest of the parameters to their
reference values and increasing  $n_0$ (or equivalently the Thomson optical
depth along the axis of the jet $\tau_{\parallel}$) makes the emergent spectra
harder (the photons are scattered more often and gain more energy from the
energetic electrons of the jet). Similarly, by increasing $R_0$, the spectrum
hardens. This is because a larger $R_0$ translates to a larger Thomson optical
depth perpendicular to the jet axis (we kept the rest of the parameters at
their reference values)  and therefore more photon scatterings. At the same
time, time lags should also depend on the dimensions of the jet and its Thomson
depth. 

Thus, since both $R_0$ and $n_0$ affect the slope of the continuum and the
time lags, it seems reasonable to run a series of models with only $R_0$
and $n_0$ varying, while the rest of the parameters of the model are kept
fixed at their reference values.  For $R_0$, we considered the values
between 12 and 110 $r_g$ (with a step of $\Delta R_0=2$),  while, for each
$R_0$, $\tau_{\parallel}$ was left free to vary between 1 and 10, with a
step of $\Delta\tau_{\parallel}=0.2$ (we discuss the model results  in
terms of $\tau_{\parallel}$, rather than $n_0$,  because it is more
informative about  the average number of scatterings that each photon has
suffered). For each run, i.e., each pair ($R_0$, $\tau_{\parallel}$), we
derived the photon index  $\Gamma_{\rm model}$ from a fit to the emergent
spectrum between 2 and 20 keV  (i.e., with no exponential cutoff, and thus
with a slightly different $\Gamma$ from that of Eq. 5) and the average time
lag, $<t_{\rm lag, model}>$, as explained in Sect. 2.2. 

\begin{figure}
\resizebox{\hsize}{!}{\includegraphics{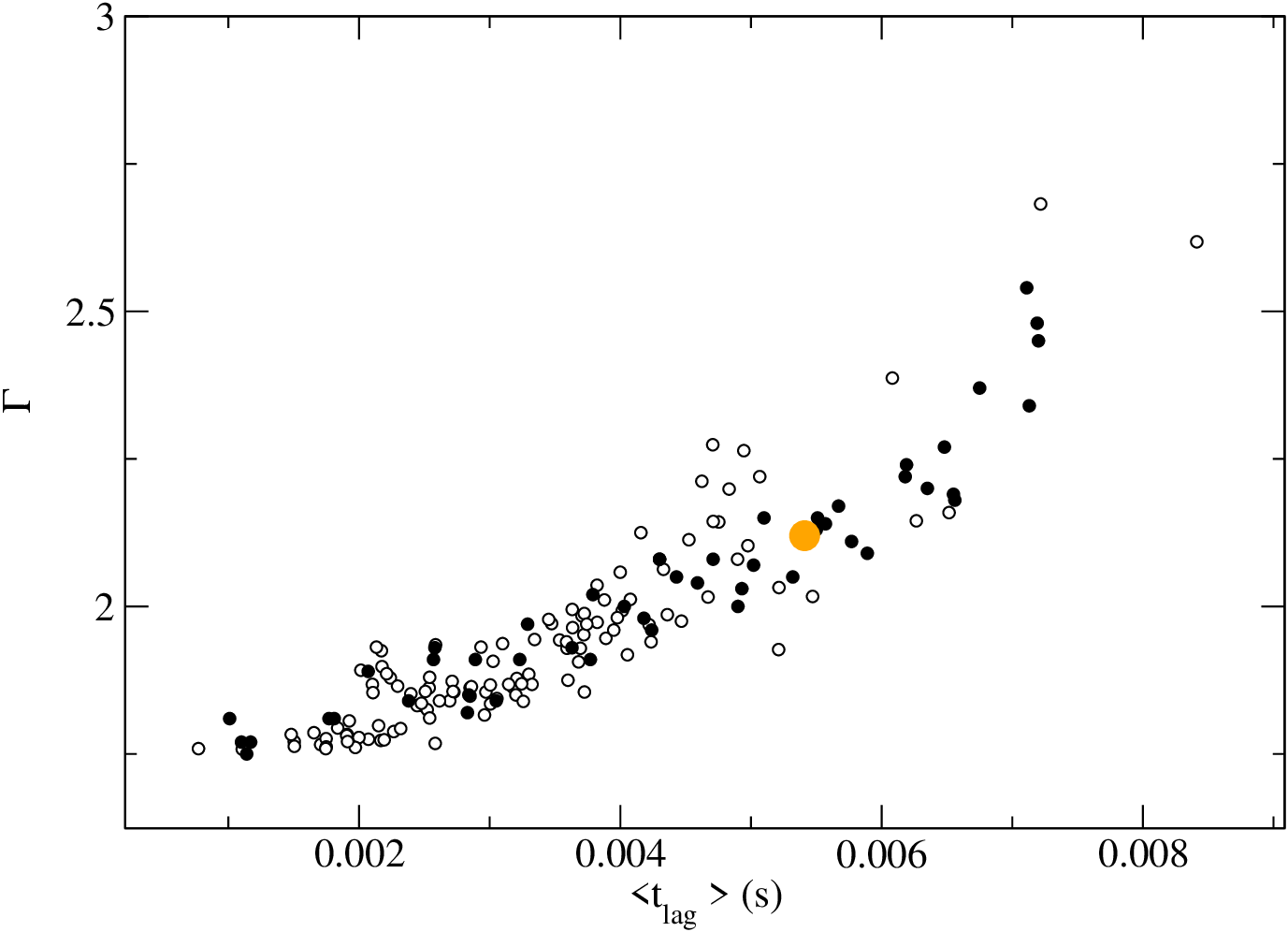}}
\caption[]{Correlation between the photon number spectral index  $\Gamma$ and
the average time lag $<t_{\rm lag}>$ between 2 - 4 keV and 8 - 14 keV in the
frequency range 3.2 - 10 Hz. Empty circles are from \citet{pott03} and
filled circles are the results of our model. The larger grey circle represents
the model with the reference values of the parameters.}
\label{gl}
\end{figure}

\begin{figure}
\resizebox{\hsize}{!}{\includegraphics{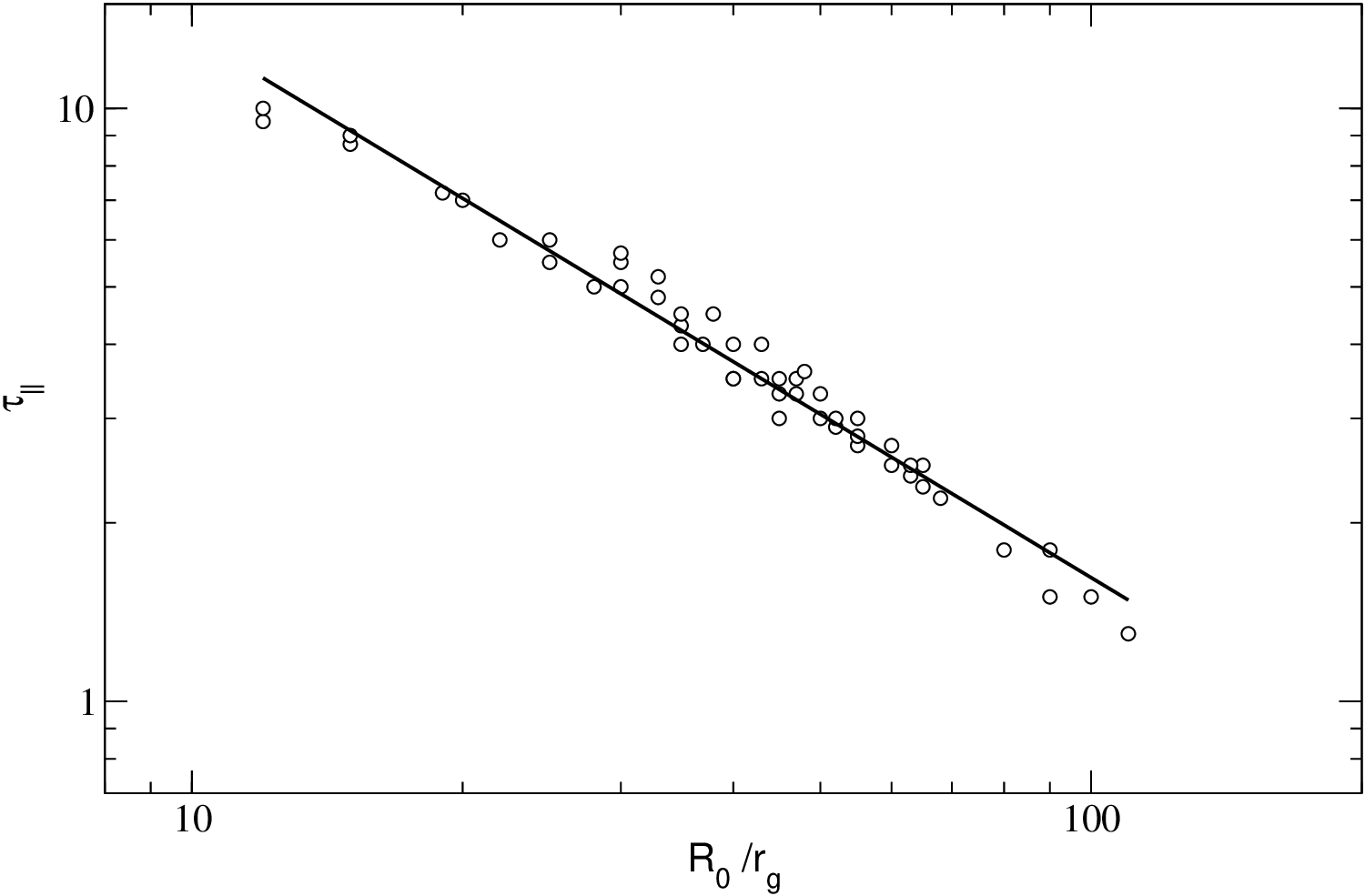}}
\caption[]{The model $R_0$ and ${\tau}_{\parallel}$ values that result in
the $\Gamma$ and $<t_{\rm lag}>$ values of Fig.~\ref{gl}, and which agree with 
the \citet{pott03} results.}
\label{glm}
\end{figure}

In Fig.~\ref{gl} we represent the $\Gamma-<t_{\rm lag}>$ relation of
\citet{pott03} (i.e. the observations), while  Fig.~\ref{glm} shows the
input model parameters ($R_0$, $\tau_{\parallel}$) that generate the points
($\Gamma_{\rm model},<t_{\rm lag, model}>$) of Fig.~\ref{gl}. We show only
those model results that agree best with the observed points.  We find that
the model can reproduce well the  observed  $\Gamma-<t_{\rm lag}>$ relation
in Cyg X-1 if the radius at the base of the jet and the optical depth along
the jet axis vary  between $R_0 \approx 5$ and 50 Schwarzschild radii,
$r_s$ ($r_s=2r_g$), and  $\tau_{\parallel} \approx 1.5-10$, respectively.

However, even more interesting is that, for the model to explain the
observational spectral index -- time lag relation, $R_0$ and $\tau_{\parallel}$
must change in a highly correlated way: as $R_0$ increases, $\tau_{\parallel}$
must decrease.  In fact, we find that a  straight-line fit to the points 
(Fig.~\ref{glm}) is very close to a $\tau_{\parallel}$ (or
$n_0$) $\propto R_0^{-1}$ relation.  Generally speaking, when $R_0$ increases,
so does the  average time lag. The spectrum should also harden at the same time.
However, if $\tau_{\parallel}$ (or $n_0$) decreases inversely proportional to
$R_0$, it affects the spectrum more and forces it to steepen exactly as
observed.  

\citet{pott03} computed $\Gamma$ and $<t_{\rm lag}>$  using data from \xte\
observations that lasted $2-7$ ks and were performed every $\sim 10$ days
over a period of 3.5 years. In effect, these values characterize the
long-term evolution of the physical characteristics of the X-ray emitting
source. The results that we present in this work show that our jet model
can explain these long-term changes if we assume reasonable variations of
just two model parameters, namely the width of the jet at its base and the 
optical depth along the jet axis. 

It is important to point out that the $n_0 \propto R_0^{-1}$  (or
$\tau_{\parallel} \propto R_0^{-1}$, if all other parameters are kept at their
reference values) relation  implies (via the continuity equation)  that
$\dot{M}_{\rm out}(R_0)\propto R_0$, i.e., that the mass ejection rate in the
jet  increases linearly with $R_0$.

\subsection{The spectral index -- Lorentzian peak frequencies relation}

We can also explain, to some extent, the ``$\Gamma$--Lorentzian peak
frequencies" relation as follows. As mentioned in the Introduction, 
$M_{\rm out}/\dot{M}_{\rm out}$ is a characteristic  timescale for ejecting
all the available mass. If the ejection process operates as a damped
oscillator with frequency proportional to the inverse of this time scale, 
i.e., frequency $= C\dot{M}_{\rm out}/M_{\rm out}$ (where $C$ is a
proportionality constant), then $n_0$  and $R_0$, and hence the emitted jet
flux, will vary with the same frequency. As a result, a Lorentzian with
frequency  $C\dot{M}_{\rm out}/M_{\rm out}= C\pi m_p
v_{\parallel}R_0^2n_0/M_{\rm out} =C^{\prime}R_0^2n_0$  will appear in the
power spectrum. Here we have assumed that $M_{\rm out}$ and 
$v_{\parallel}$ are constant.

\begin{figure}
\resizebox{\hsize}{!}{\includegraphics{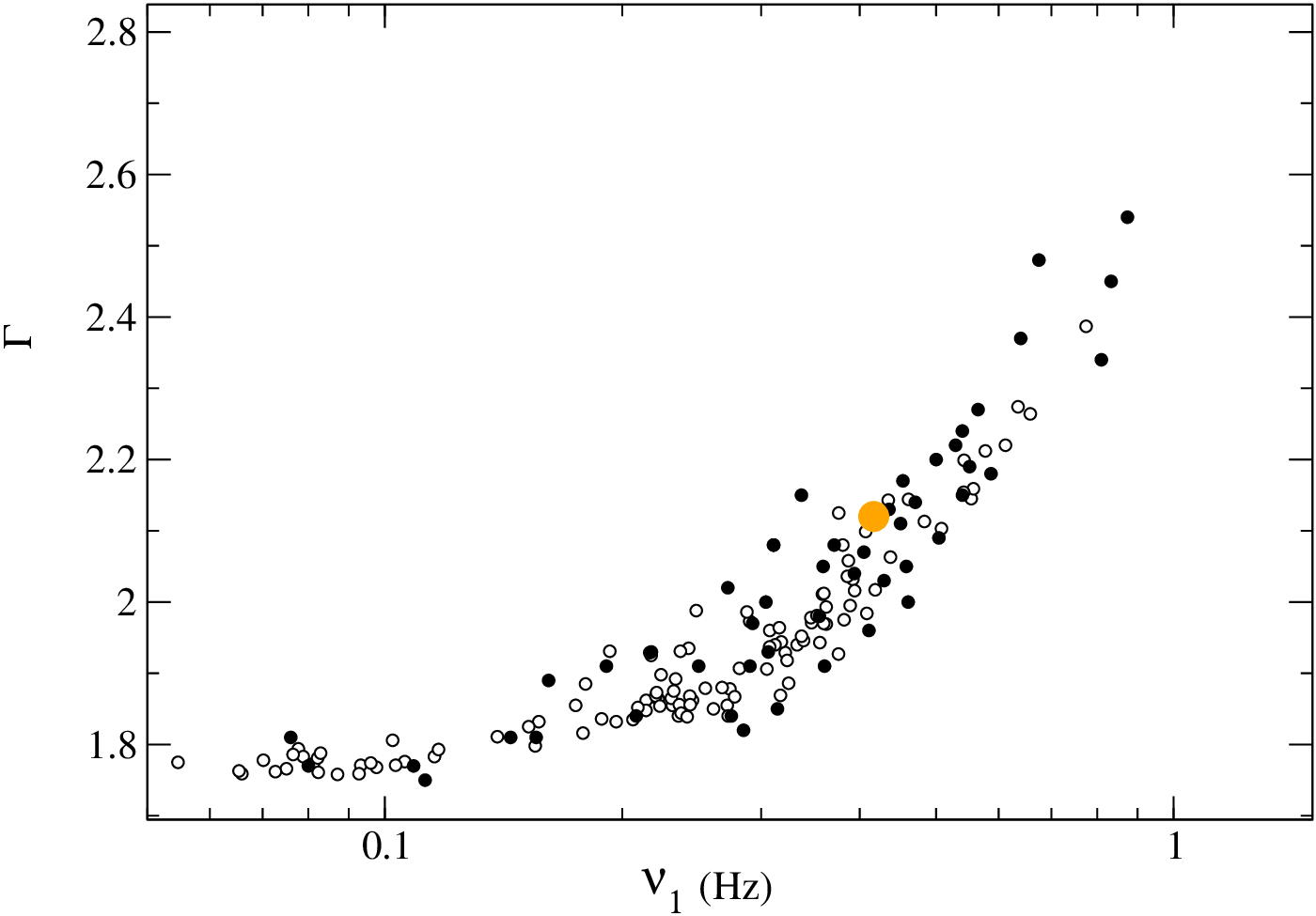}}
\caption[]{Correlation between the photon number spectral index 
$\Gamma$ and the frequency $\nu_1$ of the first Lorentzian component of
the power spectrum. Empty  circles are from \citet{pott03} and filled circles
are the results of our model. The larger grey circle represents the 
model with the reference values of the parameters.}
\label{gf}
\end{figure}

In Fig. \ref{gf} we plot $\Gamma$ versus $\nu_1$, i.e. the peak frequency
of the first Lorentzian of \citet{pott03} and the $\Gamma$ versus
$C^{\prime}R_0^2n_0$ plot that result from our simulations,  {\it using the
same $R_0$ and $n_0$ parameter values, which reproduce the
$\Gamma-<t_{\rm lag}>$ relation of \citet{pott03} } (i.e., the values
plotted in Fig. \ref{glm}) for  $C^{\prime}= 1.2 \times 10^{-33}$ cm
s$^{-1}$. Clearly, the model-predicted ``$\Gamma$--Lorentzian peak
frequency $\nu_1$" relation agrees with the data.

Here $C^{\prime}$ is one extra model parameter, whose value is difficult to
interpret at the moment (as it should be associated with the unknown
details of the ejection mechanism at the base of the jet).  This parameter
is necessary for adjusting the {\it normalization} of the
$\Gamma-\nu_1$ relation. However, according to our model, the {\it shape}
of this relation only relies on the dependence on $R_0$ and $n_0$ through
the combination $R_0^2n_0$. We believe that  the similarity between the
shape of the  observed and the model-predicted $\Gamma-\nu_1$ relation,
when $R_0$ and $n_0$ take the same values that reproduce the
$\Gamma-<t_{\rm lag}>$ relation, is an encouraging result, which
supports our model. Since the ratios of the Lorentzian peak frequencies
$\nu_i, i=1,2,3,4,$ in Cyg X-1 are constant, there is no need to reproduce
the $\Gamma-\nu_j, j=2,3,4,$ correlations. 

We note that the $\nu_i\propto R_0^2n_0$ relation, combined with the $n_0\propto
R_0^{-1}$ one (which is necessary for the model to explain the  $\Gamma$ - 
$<t_{\rm lag}>$ correlation as we showed above), implies that $\nu_i\propto
R_0$. In other words, we can explain the relation between $\Gamma$ and
Lorentzian peak frequencies only if these frequencies {\it increase} with
increasing $R_0$. This is contrary to what is usually assumed, i.e., that {\it
time scales} increase with increasing size, and hence that frequencies should
{\it decrease} with increasing $R_0$. 

In our case, the $\nu_i\propto R_0$ relation is due to the fact that we assumed
that $M_{\rm out}$ does not depend on radius.  If the mass ejection rate in the
jet increases with $R_0$ (as the model requires so that  $n_0\propto R_0^{-1}$,
and hence $\Gamma$ and  $<t_{\rm lag}>$ correlate as observed) then the time
scale needed to eject all the available mass to the jet should naturally
decrease with increasing radius.

\subsection{The $\Gamma$ -- radio flux relation}

If indeed it is $n_0$ and $R_0$ that vary with time in such a way that $n_0
\propto R_0^{-1}$, then we expect that the radio flux should also vary, even if
all the other model parameters remain constant. In fact, it is reasonable to
believe that there will be a correlation between this flux and $\Gamma$, time
lags, and Lorentzian frequencies. It is this possibility that we investigate in
this section.

The radio emission of the jet in our model is a result of the synchrotron
emission of the fast-moving electrons in the magnetic field of the jet. The
accurate calculation of the complete synchrotron spectrum can be achieved
numerically by dividing the jet in slices across the $z$-axis, and
calculating the synchrotron emission and absorption in each slice assuming
a power-law electron distribution (Paper III).  There it is shown that the
radio emission from the jet is characterized by a flat spectrum with flux
density close to observations. For the purposes of this work, we derive the
scaling of the radio flux density with the model parameters. 

The observed flux density $P_\nu$ that comes for height $z$ of the jet
scales as (e.g. Rybicki \& Lightman 1979)

$$
P_{\nu}\propto C_\alpha B^{\frac{\alpha+1}{2}}\gamma_{\rm min}^{\alpha-1}n_e
\nu^{\frac{1-\alpha}{2}}R^2(z)z,
\label{P_nu}\eqno(6)
$$
\noindent where $\gamma_{min}$ stands for the minimum Lorentz factor of the
electron distribution, and $\alpha$ for its power-law exponent (Paper III). 
The function $C_\alpha$ depends only on the index $\alpha$.
Using Eqs. (1), (3), and (4), the last expression can be rewritten as
$$
P_{\nu}\propto C_\alpha B_0^{\frac{\alpha+1}{2}}\gamma_{\rm min}^{\alpha-1}n_0
\nu^{\frac{1-\alpha}{2}}R_0^2z_0(z/z_0)^{\frac{1-\alpha}{2}}.
\eqno(7)
$$ 

The synchrotron radiation is strongly self-absorbed below a characteristic 
frequency (the turn-over frequency) $\nu_{\rm t}$. The synchrotron 
absorption coefficient is given by the expression 
(e.g. Rybicki \& Lightman 1979)
$$
a_{\nu}=A_{\alpha}n_e\gamma_{\rm min}^{\alpha-1}B^{\frac{\alpha+2}{2}}\nu^{-\frac{\alpha+4}
{2}},
\label{a_nu} \eqno(8)
$$
where $A_{\alpha}$ only depends on the index $\alpha$.
The turn-over frequency in a slice of the jet can be estimated as the frequency for which the
optical depth to synchrotron absorption becomes unity across the width of the
jet, i.e., $a_{\nu} R(z)\simeq 1$. Solving this
expression for $z_\nu/z_0$ and using Eqs. (1), (3),
(4), and (8), we have
$$
z_{\nu}/z_0=A'_{\alpha}\nu^{-\frac{\alpha+4}{\alpha+3}}n_0^{\frac{2}{\alpha+3}}
\gamma_{\rm min}^{\frac{2\alpha-2}{\alpha+3}}B_0^{\frac{\alpha+2}{\alpha+3}}
R_0^{\frac{2}{\alpha+3}}.
\eqno(9)
$$   

Most of the emission seen at frequency $\nu$ comes from the height of the jet
that corresponds to the thick-thin transition $z_\nu$.
Inserting expression (9) into expression (7) and using Eq. (2), 
we have for the
radio flux density, $P_{\nu}$, that one measures at a certain
frequency, $\nu$,  

$$
P_{\nu}\propto C'_\alpha R_0^{\frac{\alpha+7}{\alpha+3}}z_0^{\frac{\alpha-1}{\alpha+3}}
B_0^{\frac{3\alpha+5}{2\alpha+6}}\tau_{\parallel}^{\frac{4}{\alpha+3}}
\gamma_{\rm min}^{\frac{4\alpha-4}{\alpha+3}}\nu^{\frac{\alpha-1}{2\alpha+6}},
\eqno(10)
$$
where $C'_\alpha$ is a {\it decreasing} function of $\alpha$.
This is expected since a steeper electron distribution (with the rest of the
parameters fixed) leads to weaker radio emission.

If only $R_0$ and $\tau_{\parallel}$ vary with time according to the
relation  $\tau_{\parallel}\propto R_0^{-1}$ as we argued above, then the
above equation implies that $P_{\nu}\propto
R_0^{\frac{\alpha+7}{\alpha+3}}R_0^{\frac{-4}{\alpha+3}}$ or
$P_{\nu}\propto R_0$. In other words, we expect the radio flux to increase
with increasing width of the jet. Since as $R_0$
increases (and $n_0\propto R_0^{-1}$) the spectrum steepens, we should also
expect a positive correlation between $P_{\nu}$ and $\Gamma$: the radio
flux should increase as the spectrum steepens. 

\begin{figure}
\resizebox{\hsize}{!}{\includegraphics{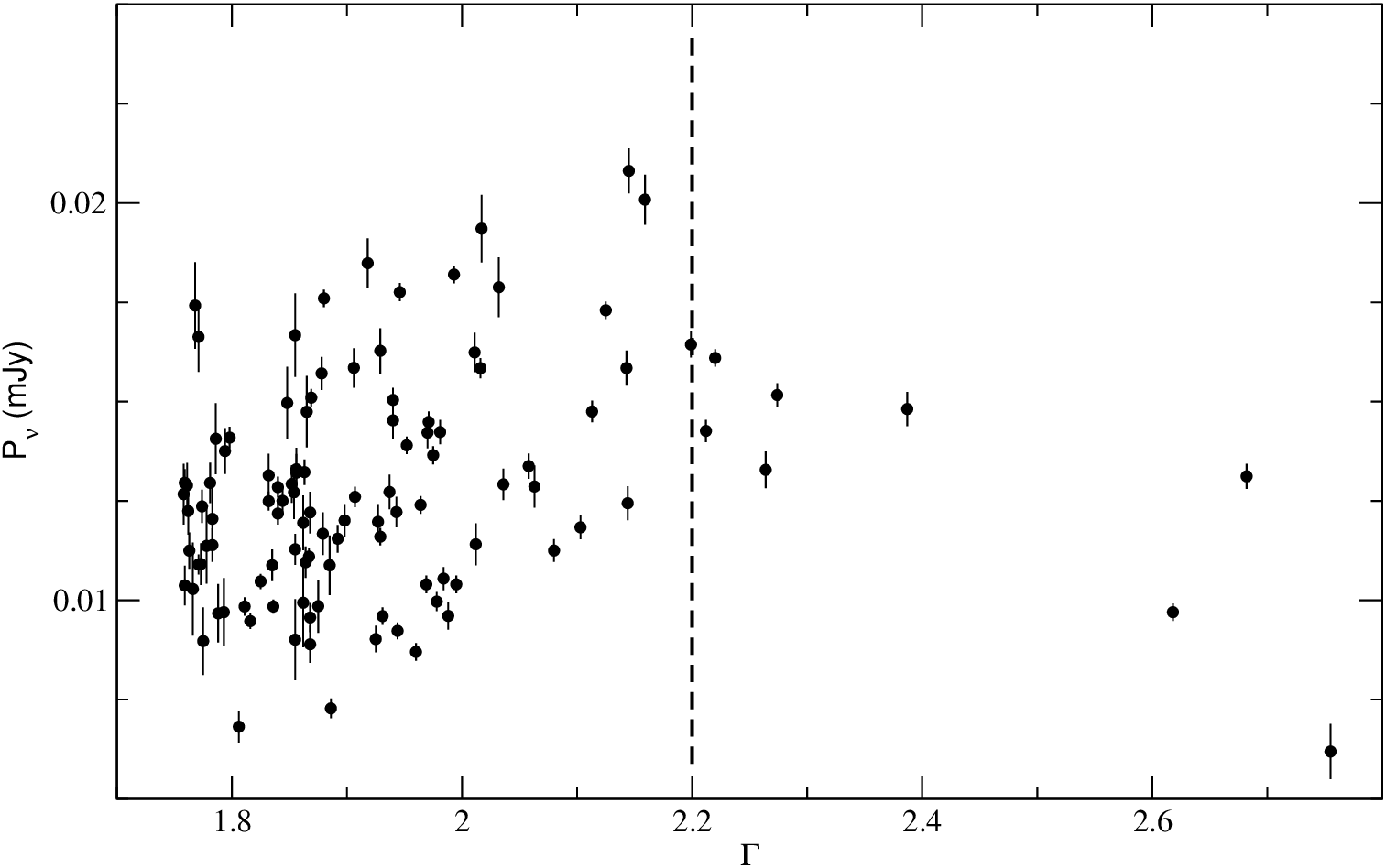}}
\caption[]{The 15 GHz flux vs the photon number spectral index, 
$\Gamma$, of Cyg X-1 between 1998 and 2001.}
\label{rg}
\end{figure}

In Fig.~\ref{rg} we plot the average 15 GHz radio flux, $P_{15\rm GHz}$, as
observed with the Ryle telescope, binned over a period of $\pm 2$ days
before and after the time of the \xte\ observations that \citet{pott03}
used to determine $\Gamma$.  (The results do not change significantly if we
bin the radio light curve using a  smaller or a larger bin size.)   One
can see that there is a general trend for  $P_{15\rm GHz}$ to increase with
increasing $\Gamma$, as expected, until $\Gamma$ reaches the value of $\sim
2.2$. When we use the  $(P_{15\rm GHz}, \Gamma<2.2)$ data points to
compute Kendall's $\tau$, we find that $\tau=0.21$. The probability of 
obtaining  this value in the case of no association between $P_{15\rm GHz}$
and $\Gamma$ is less than 0.2\%. We conclude that, as long as $\Gamma<2.2$,
there is a  weak, but statistically significant, correlation between
$P_{15\rm GHz}$ and $\Gamma$ in the sense that,  on average, the radio
flux increases as the spectrum steepens. 

According to Eq. (10), the break of the $P_{15\rm GHz} - \Gamma$  
correlation at $\Gamma \approx 2.2$, as the source moves into
intermediate/transition states (and then to the thermal state),  suggests
that, as $R_0$ keeps increasing, then $B_0$  may decrease. Alternatively,
the transition may be characterized by weaker particle acceleration leading
to either lower values of $\gamma_{\rm min}$ or higher values of the
electron index $\alpha$. If real, this result  offers a hint as to what
happens as the source  approaches the thermal state: the width of the jet
increases and the magnetic field diminishes (which explains the silence in
radio) and there is weaker particle acceleration taking place in  the jet
(which  explains the weaker X-ray power-law emission).

\section{Summary and conclusions}

In a series of three papers (Papers I, II, and III) it has been shown that our
jet model can explain a number of observations regarding black-hole X-ray
binaries in the hard state.  These are: a) the energy spectrum from radio to
hard X-rays;  b) the time-lag of the hard photons relative  to the softer ones
as a function of Fourier frequency; c) the flattening of the power spectra  at
high frequencies with increasing photon energy and the narrowing  of the
autocorrelation function. 

In this paper we investigate whether our model can explain the long term
variations in the spectral and timing properties of Cyg X-1, which are
closely linked. The main result of our work is that we can reproduce the
$\Gamma-<t_{\rm lag}>$ correlation of \citet{pott03}  if we assume
reasonable variations in just two  model parameters around their mean
values: the  width of the jet at its base and the optical depth along the
jet's axis (or equivalently the electron density at the base of the jet).
We find that, if $\tau_{\parallel}\approx 1.5-10$, $R_0\approx 5-50 r_s$ and
$\tau_{\parallel}\propto R_0^{-1}$, then the model predicts a
$\Gamma-<t_{\rm lag}>$ relation that is almost  identical to the one that
\citet{pott03} observed. Furthermore, if the flux variability operates on a
time scale which is proportional to 
$M_{\rm out}/\dot{M}_{\rm out}$ (i.e. the
characteristic time needed to eject all the available matter to the jet),
then the same $\tau$ and $R_0$ variations can also explain the {\it
shape} of the ``spectral index - Lorentzian peak frequencies" correlation
of \citet{pott03}. Finally, given the model parameter variations that are
needed to explain the observed "spectral - timing" correlations, the
model predicts a "radio flux -- $\Gamma$" correlation, 
which, as we show, does indeed hold in Cyg X-1. 

The $\tau_{\parallel}\propto R_0^{-1}$ relation, which is  necessary to
explain  the observed  $\Gamma-<t_{\rm lag}>$ relation, is a  natural
outcome of the model if simply $\dot{M}_{\rm out}\propto R_0$. We propose
that this is the underlying ``fundamental" relation that governs the  {\it
long-term} evolution of the physical characteristics of the jet,  hence of
the observational spectral and timing properties of the source. The
physical reason for this relation is unclear at the moment; but whatever
its origin, the relation itself does make sense because as the rate of the
mass ejected in the jet increases, it seems reasonable to assume that the
jet will become ``bigger", i.e. $R_0$ will also increase. In fact, in this
way, one can explain reasonably well all the long-term observed changes 
with variations in one fundamental physical parameter of the system, namely
the accretion rate. As the accretion rate increases, $\dot{M}_{\rm out}$
should also increase. If the jet size increases proportionally to
$\dot{M}_{\rm out}$, then $n_0 \propto R_0^{-1}$, and  hence the $\Gamma$
-- $<t_{\rm lag}>$ relation.  We hope that this ``accretion rate --
ejection rate -- jet size"  relation will provide an important and useful
hint for all models that try to explain the connection between accretion
flows and outflows, hence the formation of jets in regions close to
the central engines in accreting compact objects. 

We find that the $\dot{M}_{\rm out}\propto R_0$  relation predicts that the
radio flux should increase linearly with the photon number spectral index.
To the best of our knowledge, such a relation has never been suggested or
even detected either in Cyg X-1 or in any other GBH in the past.  We show
that a linear  ``radio flux -- spectral index" relation does exist  on
long time scales in Cyg X-1. We believe that this result strongly supports
our view that a) X-rays in Cyg X-1 are produced by Compton upscattering in
the jet of soft photons from the accretion disk and b) the jet's size and
optical depth evolve with time because $\dot{M}_{\rm out}\propto R_0$. 
Furthermore,  that the ``radio flux --  spectral index" relation
breaks at index values higher than $\sim 2.2$,  provides interesting hints
into what may be happening as the source moves to the intermediate and
thermal state. Most probably the magnetic field  decreases and/or there is
weaker particle acceleration in the jet (i.e. the electron distribution
index, $\alpha$, becomes larger)   as the accretion rate increases (as
inferred from the increase in the disk thermal emission).

The explanation of the short time-scale variability properties of Cyg X-1
is significantly more challenging. Our model cannot explain why  four
Lorentzian components exist in the source's power spectrum (assuming of
course that the Lorentzians are the fundamental building blocks of the
power spectrum in Cyg X-1 and other GBHs). However, it can explain, to some
extent, the way they vary with the source's photon  number spectral index.
If indeed the long-term evolution of the source is governed by variations
in $R_0$ and $\dot{M}_{\rm out}$, it is natural to assume that the
short-term variations are caused by short-term, random variations of 
$\dot{M}_{\rm out}$ (and hence of $n_0$ or $\tau_{\parallel}$, and finally
X-ray flux), around its mean value. We find that, if these variations
operate on a time scale that is proportional to the characteristic  time
needed to eject all the available matter to the jet, then the model can
explain the $\it shape$ of the ``$\Gamma$ - Lorentzian peak frequencies"
relations of \citet{pott03}. There is no reason why the model  ``$\Gamma -$
inverse time needed to eject the  mass to the jet" relation should have the
same shape as the observed ``$\Gamma -$ Lorentzian peak frequencies". Thus
we do not find this agreement to be coincidental. Instead, it is one more
evidence in favor of the jet model and the parameters' evolution we
present in this work, and we find it  operates on long time scales in Cyg
X-1. 

Our results suggest that the characteristic  frequencies seen in the power
spectrum should increase with increasing size of the source, i.e. $R_0$, a
result that goes against current views. However, the characteristic
frequencies may not scale directly with the source size, but with accretion
rate, hence  $\dot{M}_{\rm out}$.  As argued above, it seems reasonable to
think that, as the accretion rate increases, so too the mass ejection rate
to the jet. Perhaps, then, it is the accretion rate (through $\dot{M}_{\rm
out}$)  that drives the variations in both $R_0$ and on the characteristic
time scales. A  physical explanation as to how this may happen and as to
the value (and number) of characteristic time scales, requires knowledge of
the details of the acceleration/ejection process of the jet, and is beyond
the scope of the present work. However, as we mentioned above, we hope that
the results we present here will help the investigation of the proper
mechanism responsible for the launch of the jets and winds in accreting
compact objects. 

We note that our conclusion, presented in Sect. 3.2, that the
Lorentzian peak frequencies $\nu_i$ {\it increase} with the size $R_0$ of
the base of the jet, holds for a given source (in our case Cyg X-1).  If
one wanted to extend our model to other sources, then the mass of the black
hole comes in.  The argument goes as follows:  Making the reasonable
assumption that $M_{\rm out} \propto R_0^2$ and using $n_0 \propto
R_0^{-1}$, we find that the $\nu_i$ scale as $1/R_0$.  Since $R_0$ is
typically a few times $r_g$, we conclude that {\it the Lorentzian peak
frequencies $\nu_i$ are inversely proportional to the mass of the black
hole}.

In closing, we want to point out that the above results and conclusions are
consistent with the fact that, in the thermal state of Cyg X-1, the time
lags versus Fourier frequency are {\it identical} to those in the hard
state \citep{pott00}.  In the thermal state, the radio emission is very
weak and possibly undetectable,  {\it but we think that the jet is there}
and that it produces, by Comptonization,  both the time lags versus Fourier 
frequency and the steep power-law energy spectrum above $\sim 10$ keV.
According to our model, the suppression of both the radio and the X-ray 
emission in this state should be caused by a magnetic field suppression 
and a weaker particle acceleration, respectively.

\begin{acknowledgements}
We thank K. Pottschmidt for sending us the Cyg X--1 time lag data. 
This work was supported in part by the European Union Marie Curie grant 
MTKD-CT-2006-039965. 
\end{acknowledgements}

\end{document}